\documentstyle[12pt,psfig]{article}
\begin{document}
\newcommand{\scr}{\sin^2 \hat{\theta}_W (m_Z)}
\newcommand{\slept}{\sin^2 \theta_{eff}^{lept}}
\newcommand{\smallw}{{\scriptscriptstyle W}}
\newcommand{\smallz}{{\scriptscriptstyle Z}}
\newcommand{\sef}{\sin^2 \theta_\smallw^{eff}}
\newcommand{\msbar}{\rm{\overline{MS}}}
\newcommand{\be}{\begin{eqnarray}}
\newcommand{\en}{\end{eqnarray}}
\newcommand{\kcar}{\hat{k}}
\newcommand{\scar}{\hat{s}}
\newcommand{\cc}{\hat{c}}
\newcommand{\alc}{\hat{\alpha}}
\newcommand{\nn}{\noindent}
\newcommand{\ew}{electroweak~}
\newcommand{\mz}{m_\smallz}
\newcommand{\mw}{m_\smallw}
\nn     \hspace*{10.6cm} NYU--TH--94/04/01\\
\hspace*{11.9cm} April 1994\\

\vspace*{2cm}

\centerline{\large{\bf Evidence for Bosonic Electroweak Corrections }}
\centerline{\large{\bf in the Standard Model$^*$}}

\vspace*{.8cm}

\centerline{\sc Paolo Gambino }

\vspace*{.4cm}

\centerline{ Department of Physics, New York University, 4 Washington
Place,}
\centerline{ New York, NY 10003, USA.}
\vspace*{.8cm}
\centerline{\sc Alberto Sirlin$^{**}$ }

\vspace*{.4cm}

\centerline{ Department of Physics, Brookhaven National Laboratory,}
\centerline{ Upton, Long Island, NY 11973, USA.}

\vspace*{1cm}

\begin{center}
\parbox{14cm}
{\begin{center} ABSTRACT \end{center}
\vspace*{0.15cm}
We present strong indirect evidence for the contribution of bosonic \ew
corrections in the Standard Model. Although important conceptually,
these corrections give subleading contributions in current high energy
experiments, and it was previously thought that they are difficult to
detect. We also discuss the separate  contribution of the Higgs boson.}
\end{center}
\vspace*{1.9cm}

\nn{\small $ ^*$ This work was supported in part by the High
 Energy Theory Physics
program of the U.S. Department of Energy under Contract No. DE-AC02-76CH00016
and in part by  NSF, Grant No. PHY-9313781.

\nn $^{**}$ Permanent address:  Dept. of Physics, New York University,
 4 Washington Place, New York, NY 10003, USA }
\vfill
\newpage
\nn It has been known for a long time that radiative corrections
 play an important role in the analysis of the Standard Model (SM)
predictions. For example, if their contribution at low energies were not
included, there would be a very large violation of the unitarity of the CKM
matrix and the SM would be placed in severe jeopardy \cite{1}.
At high energies, the current estimates on $m_t$ are derived from
\ew radiative
corrections. However, unlike the problem of universality,
where the corrections involve virtual $W^{\pm},\gamma$, and $Z^0$\cite{2},
the dominant corrections in current high-energy experiments
are fermionic in nature. That is, at one-loop, they involve only virtual
fermions. They are responsible, for instance for the large logarithms
associated with the running of $\alpha$\cite{3,4} and for the sizable
contributions of the $t-b$ sector, from which the $m_t$ estimates are derived.
An exception to this statement is the $Z^0\to b \bar b$ vertex, where
 there are also large corrections involving virtual bosons.

However, aside from the fermionic contributions, there are also, at high
energies, conceptually very important bosonic corrections mediated by
virtual $W^\pm, \gamma, Z^0$, and $H$. They involve the plethora of
bosonic couplings of the SM, including the well-known tri-linear
vertices, and affect self-energies, vertex, and box diagrams
(in four fermion processes, bosonic vertex and box diagrams
also include  virtual fermions). We recall that
this separation into fermionic and bosonic one-loop corrections
is gauge-invariant and
goes back to the early papers \cite{5}. The difficulty is that, with the
exception of the  $Z^0\to b \bar b$ vertex mentioned before, in current high
energy experiments the bosonic corrections give subleading contributions,
significantly smaller than their fermionic counterparts.

The purpose of this paper is to show that the accuracy currently reached
is such that strong indirect evidence for the presence in the SM of these
important subleading corrections  can be obtained. Specifically,
our approach consists in removing the full block of bosonic \ew
corrections in two highly
precise determinations of the $\msbar$-parameter $\scr\equiv\scar^2$
and showing that, for $m_t\ge 131$ GeV (the current lower bound), this
leads to a sharp disagreement. In the analysis all the fermionic
contributions are retained.

One precise determination of $\scar^2$ can be readily
derived from the result $\sef=0.2316\pm 0.0004$, obtained
from the combination
of LEP and SLC asymmetry values \cite{6}. Here $\sef$, also called
$\slept$, is the effective  mixing parameter in the $Z^0\to \ell \bar\ell$
on-shell amplitude.
It has been recently shown that $\sef=Re \hat{k}_\ell(\mz^2)\ \scar^2$,
where $\hat{k}_\ell(\mz^2)= 1 + {\cal O}(\alpha)$
represents the relevant radiative corrections \cite{7}. Removing the bosonic
contributions, one obtains
$Re \hat{k}_\ell (\mz^2)= 1.0060$, so that
\be
(\scar^2)_{tr}=0.2302\pm.0004 \ \ \ \ \ \ (Asymmetries).
\label{as}\en
Henceforth the subscript {\it tr} reminds us that this value corresponds
to a ``truncated'' version of theory, with bosonic contributions
removed in the \ew corrections. In evaluating
$Re \hat{k}_\ell (\mz^2)$ we have neglected small ${\cal O}(\alpha
\hat{\alpha}_s)$ corrections that were retained in Ref.\cite{7}.

A second precise determination  can be obtained from $G_\mu, \alpha, \mz$:
\be
(\scar^2)_{tr} (\hat{c}^2)_{tr}= \frac{A^2}{\mz^2 (1-(\Delta\hat{r})_{tr})},
\label{dr}\en
where $A^2=\pi \alpha/\sqrt{2} G_\mu$ and $\Delta\hat{r}$ is a basic
correction of the electroweak theory \cite{8}, from which we have removed
once more the bosonic contributions. The comparison between the two
determinations of $(\scar^2)_{tr}$, from Eqs.\ref{as} and \ref{dr},
is shown in Fig.1, as a function of $m_t$.
In evaluating $(\Delta\hat{r})_{tr}$,  we employed the recent preliminary
values $\mz=91.1899\pm 0.0044$GeV \cite{9}, and $e^2 Re[\Pi^{(5)}_{
\gamma\gamma}(0)
-\Pi_{\gamma\gamma}^{(5)}(\mz^2)]=0.00283\pm 0.0007$
for the five flavor contribution to the photon vacuum polarization
function \cite{10}.
We have neglected small ${\cal O}(\alc^2)$  and ${\cal O}(\alc\alc_s)$
corrections,
except for the leading two-loop effects of ${\cal O}(\alc_s G_\mu m_t^2)$,
which we have incorporated using a simple method recently discussed
by one of us \cite{11} and $\alc_s(\mz)=0.12$; we have also
removed the irreducible two-loop effects of ${\cal O}(\alpha^2 m_t^4)$
\cite{12} as they involve virtual bosons.

It is apparent from Fig.1 that the removal
of the bosonic component of the \ew
corrections leads to a sharp disagreement. At the lower bound $m_t=131$ GeV,
 the value obtained from Eq.\ref{dr} is
\be
(\scar^2)_{tr}= 0.2282\pm 0.0003 \ \ \ \ \ \
\ \ (G_\mu,\alpha,\mz,m_t=131{\rm GeV})
\en
and we see that the difference with Eq.\ref{as} amounts to 4$\sigma$ (if the
recent SLC value were not included,  the top curve in Fig.1 would be shifted
upwards by $\approx 0.0006$, the error would be slightly increased, and the
difference would be 4.5$\sigma$). As shown in Fig.1,
the discrepancy rapidly increases with $m_t$. For instance, it is $7\sigma$
for $m_t=180$ GeV.
Fig.2 shows the same comparison, but with the bosonic \ew corrections
restored, in the case $m_H=300$ GeV. As expected, there is now consistency
between
the two determinations for a restricted range of $m_t $ values.
Comparing Fig.1 and 2, we see that the removal of the bosonic corrections
leads to lower values of $\scar^2$. The discrepancy arises because the effect
is much more pronounced in the ($G_\mu,\alpha,\mz$) determination.

Given the sharpness of the signal, it is natural to ask whether one can
use the same approach to probe specific components of the bosonic
corrections. For instance, can one search for signals of Higgs boson
contribution by removing them from the \ew corrections, retaining
the rest?
As $H$ does not contribute, at  one-loop level, to
$\hat{k}_\ell(\mz^2)$, the value of $\scar^2 $ derived in this case from
$\sef$ is that of the full SM \cite{7}:
\be
\scar^2=0.2313\pm 0.0004 \ \  \ \ \ \ \ (Asymmetries),
\label{full}\en
instead of Eq.\ref{as}. In order to be physically meaningful,
the removal of the Higgs contribution from $\Delta\hat{r}$ must be done in a
gauge-invariant and finite manner. Fortunately, in the SM the diagrams
involving $H$ in the self-energies contributing to $\Delta\hat{r}$
form a gauge-invariant subset.
On the other hand, they are divergent.
Therefore, one must specify the renormalization prescription and the scale
at which these partial contributions are evaluated. As $\scar^2$ is
 the $\msbar$ parameter and the \ew data are  dominated by information
at the $Z^0$-peak, it is natural to subtract the $\msbar$-renormalized
Higgs boson contribution evaluated at the $\mz$ scale.
 Neglecting non-leading ${\cal{O}}(\alpha^2)$
terms, the latter is given by
\be
(\Delta\hat{r})_{H.B.}=
 \frac{\alpha}{4\pi\scar^2}\left\{
\frac{1}{\cc^2} H(\xi)-{3\over4} \frac{\xi\ln\xi-\cc^2\ln \cc^2}{\xi-\cc^2}+
{19\over24}+\frac{\scar^2}{6\cc^2}\right\}_{\msbar}
\label{higgs}\en
where $\xi=m_H^2/\mz^2$,  $H(\xi)$ is a function studied
in Ref.\cite{5},
and the subscript $\msbar$ reminds us that
the $\msbar$ renormalization has been carried out and the scale
$\mu=\mz$  chosen.
The need to specify the scale in defining the Higgs boson contribution
can be most easily understood in the on-shell method of renormalization
\cite{5}, where one employs $\sin^2\theta_W= 1- \mw^2/\mz^2$ instead of
$\scar^2$. In that case, the relevant radiative correction is $\Delta r$,
rather than $\Delta \hat{r}$. Although $\Delta r$ is a physical observable
and is therefore $\mu$-independent, the Higgs-boson contribution is
$\mu$-dependent. Thus, a specification of the scale is necessary in its
definition.

Subtracting then the Higgs boson contribution one obtains
a new truncated version of $\Delta\hat{r}$, independent of $m_H$, from which
we can compute the corresponding $(\scar^2)_{tr}$ via Eq.\ref{dr}.
The comparison with Eq.\ref{full} is given in Fig.3.
In contrast with Fig.1, where the complete block of
bosonic contributions was
removed, there are no signals of inconsistency. This is easily understood by
noting that $(\Delta\hat{r})_{H.B.}$ vanishes for $m_H\approx 113 $ GeV.
Thus, the subtraction of Eq.\ref{higgs} is equivalent to a SM model
 calculation
with a relatively light Higgs scalar, $m_H\approx 113 $ GeV, and this is
consistent with current \ew data.

On the other hand, for $m_H=1$ TeV, $(\Delta\hat{r})_{H.B.}\approx
3.4\times 10^{-3}$, which is not far from the value
$4.0\times 10^{-3}$ derived from the asymptotic formula
$ (\Delta\hat{r})_{H.B.}\sim (\alpha/4\pi \scar^2 \cc^2)  [5/6 -
 3\cc^2/4 ]\ln (m_H^2/\mz^2)$.
As a consequence, in the $m_H=1$ TeV case,
there is an additional contribution
that raises the $\mz$ determination of $\scar^2$ in Fig.3 by $\approx
1.2\times
10^{-3}$ and therefore favors a larger $m_t$ value.
 The parameter $m_H$ occurring in the asymptotic formulae has  been
interpreted as playing the role of a regulator in
nonlinear $\sigma$-models that
describe the heavy-Higgs-boson limit of the SM \cite{long}.

In  summary, we have presented strong indirect evidence for the presence
in the SM of
bosonic \ew corrections (Fig.1). If one probes just the Higgs component
of these corrections, no evidence has been uncovered
in our very simple analysis. However, it is
likely that the signals will become sharper as the precision
increases and $m_t$ is measured. It is also likely that more
 precise evidence
for the bosonic \ew corrections and their components will emerge
if the approach we have proposed, namely their removal from the
relevant corrections, is extended systematically to global analyses.

\subsection*{Aknowledgments.}
\vskip .4cm
We would like to thank
Sergio Fanchiotti, William Marciano, and Andrea Pelissetto for very useful
discussions.

\newpage
\centerline{\psfig{figure=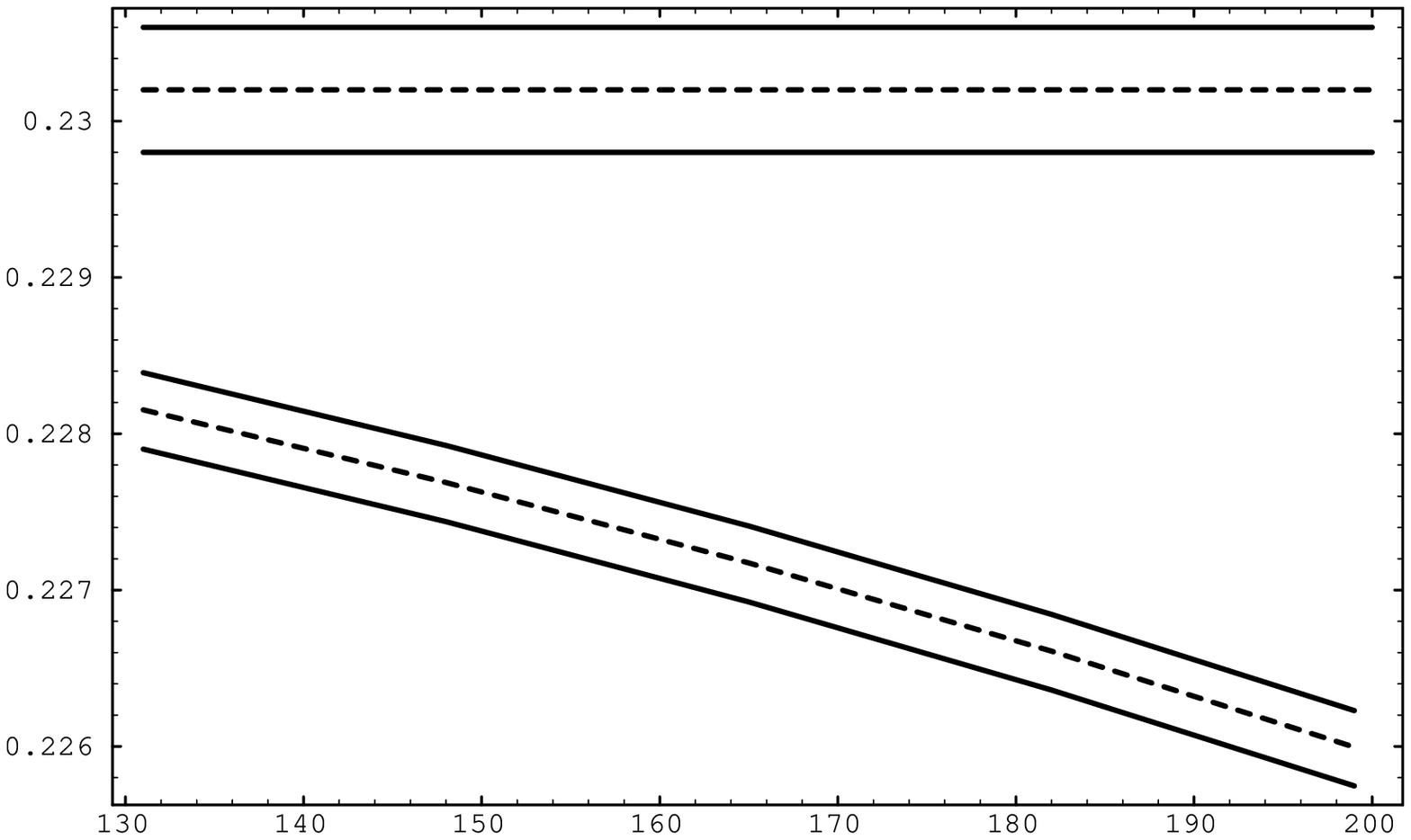,height=18.2cm}}
\centerline{\small{\bf Fig.~1} Determination of $\scr$ from asymmetries
(horizontal lines)and $\mz$ }
\centerline{\small  (bottom curves)
with the   bosonic contributions removed, as a function of $m_t$ (GeV).}
\centerline{\small The one $\sigma$ errors are indicated.}
\par
\centerline{\psfig{figure=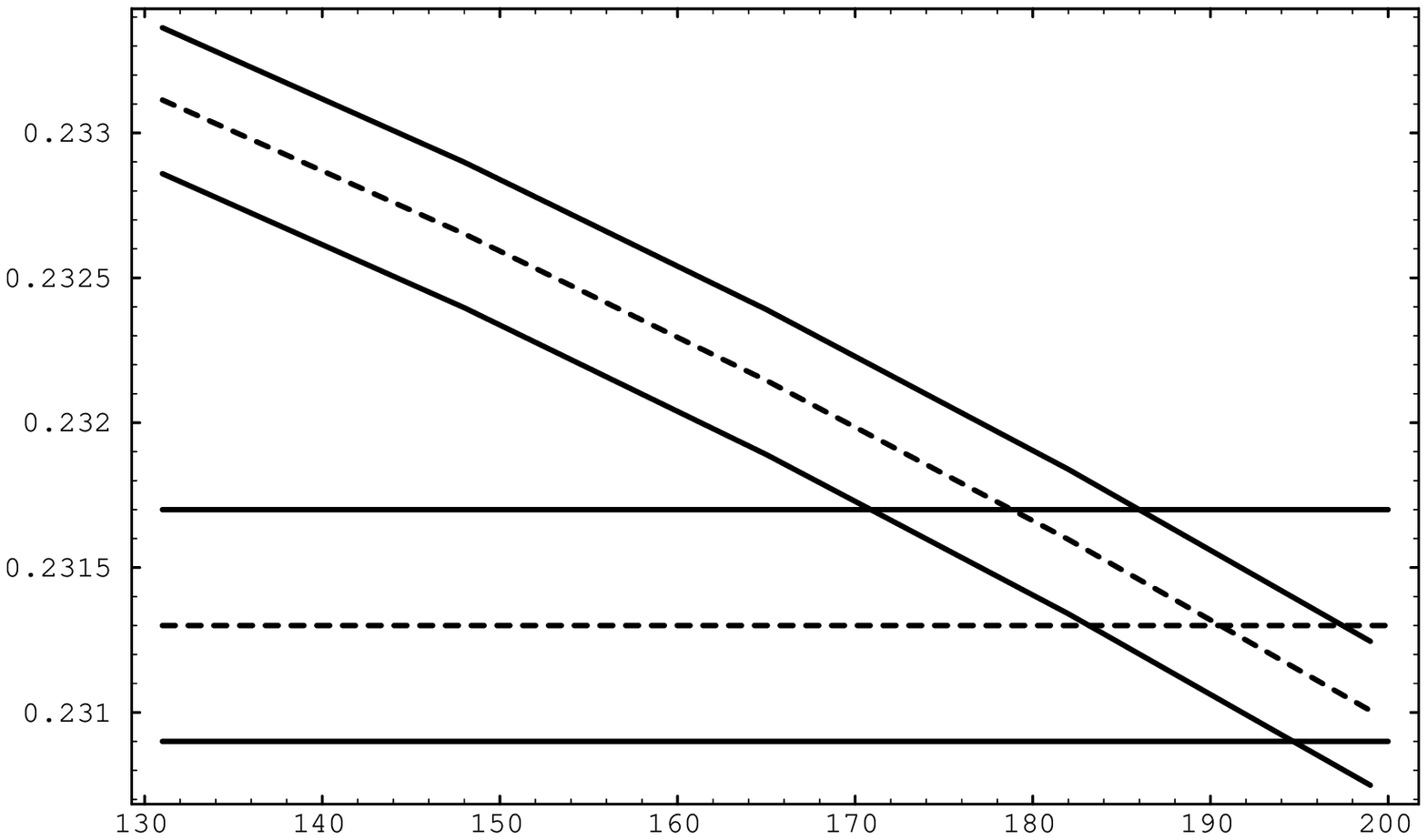,height=18.5cm}}
\centerline{\small{\bf Fig.~2} Determination of $\scr$ from asymmetries and
$\mz$ in the full SM }
\centerline{\small for $m_H=300$ GeV, as a function of $m_t$ (GeV).}
\centerline{\psfig{figure=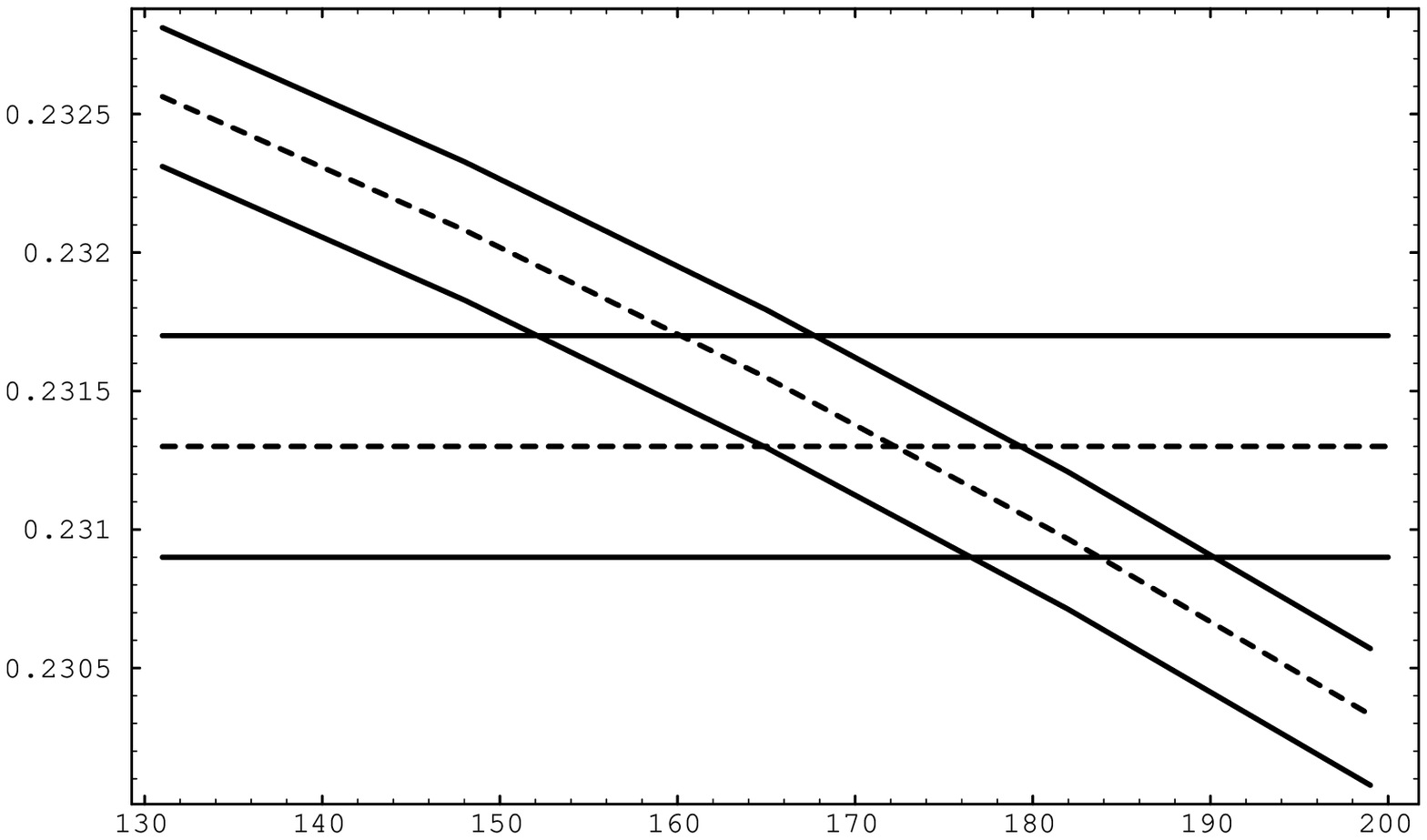,height=18.5cm}}
\centerline{\small{\bf Fig.~3} Determination of $\scr$ from asymmetries and
$\mz$ with the Higgs}
\centerline{\small  scalar contribution removed (see text),
 as a function of $m_t$ (GeV).}

\begin{thebibliography}{99}
\bibitem{1} See, for example, A. Sirlin, Phys. Rev. Lett.  \underline{72},
1786 (1994) and references cited therein.
\bibitem{2} A. Sirlin, Rev. Mod. Phys. \underline{50}, 573 (1978)
and references cited therein.
\bibitem{3} W.J. Marciano, Phys. Rev. D \underline{20}, 274 (1979).
\bibitem{4} The direct and indirect evidence
for the corrections not contained
in the running of $\alpha$ at the $\mz$ scale has been recently discussed
in Ref.\cite{1}. If the new SLC results are included,
the indirect evidence becomes much sharper.
\bibitem{5} A. Sirlin, Phys. Rev. D \underline{22}, 971 (1980);
W.J. Marciano and  A. Sirlin, Phys. Rev. D \underline{22},
2695 (1980).

\bibitem{6}B. Pietrzyk, talk given at the XXIX Rencontre de Moriond,
Savoie, France (1994).

\bibitem{7} P. Gambino and A. Sirlin, Phys. Rev. D \underline{49},
R1160 (1994).

\bibitem{8} A. Sirlin, Phys. Lett. B \underline{232}, 123 (1989);
G. Degrassi, S. Fanchiotti, and A. Sirlin, Nucl. Phys. B
\underline{351}, 49 (1991);
 S. Fanchiotti, B.A. Kniehl, and A. Sirlin, Phys. Rev. D
\underline{48}, 307,  (1993).

\bibitem{9} P. Clarke,  talk given at the XXIX Rencontre de Moriond,
Savoie, France (1994).

\bibitem{10} B.A. Kniehl, DESY Report No. 93-150 (November 1993),
to appear in {\it Proceedings of the International Europhysics
Conference on
High Energy Physics}, Marseille, France, July 22-28, 1993, edited by
J. Carr and M. Perrottet (Editions Frontiers, Gif-sur-Yvette,1994).

\bibitem{11} A. Sirlin, BNL--60161 report, March 1994.

\bibitem{12} R. Barbieri et al., Phys. Lett. B \underline{288}, 95 (1992);
J. Fleischer , O.V. Tarasov, and F. Jegerlehner, Phys. Lett. B
\underline{319}, 249 (1993);
G. Degrassi, S. Fanchiotti, and P. Gambino, CERN-TH.7180/94.


\bibitem{long} A.C.  Longhitano, Phys. Rev. D \underline{22}, 1166 (1980).
\end{thebibliography}
\end{document}